\documentclass[%
reprint,
superscriptaddress,
nofootinbib,
amsmath,amssymb,
aps,
pra,
floatfix,twocolumn
]{revtex4-1}

\usepackage{mathtools}
\usepackage{amsmath}
\usepackage{amsfonts}
\usepackage{mathrsfs}  
\usepackage{amssymb}
\usepackage{dsfont}
\usepackage{pgfplots}
\usepackage{layouts}
\usepackage{ulem}
\usepackage[makeroom]{cancel}

\usepackage{mleftright} \mleftright 
\usepackage{scalerel,stackengine}
\usepackage{color}
\usepackage{tensor}
\usepackage{braket}
\usepackage{xifthen} % for if then else
\usepackage{xspace} % for \ie and \ig
\usepackage[section]{placeins} % for /floatbarrier
\usepackage{graphicx} % include graphics
\usepackage[colorlinks=true, linkcolor=blue, citecolor=green]{hyperref}
\hypersetup{hidelinks} % Do not mark links in the pdf.
\hypersetup{bookmarksopen=true, bookmarksopenlevel=2} % Add PDF bookmarks/table of contents usable in a pdf reader, and open them until the section level.
\hypersetup{pdfauthor={}}
\hypersetup{pdfsubject={}}
\hypersetup{pdfkeywords={}}
\hypersetup{pdflang=en} % This document is written in english.
\hypersetup{pdfdisplaydoctitle=true} % In pdf viewers, display the pdftitle of the document, not the filename.
\hypersetup{pdfduplex=DuplexFlipLongEdge} % This document is printed duplex by default.
\newcommand{\vk}[3]{\tensor*[#1]{\ket{#2}}{#3}}
\newcommand{\vb}[3]{\tensor*[#1]{\bra{#2}}{#3}}
\newcommand{\op}[3]{\tensor*[#1]{\hat{#2}}{#3}}

\makeatletter

\newcommand{\Rmnum}[1]{\expandafter\@slowromancap\romannumeral #1@}
\makeatother
\newcommand*{\ie}{i.e.\@\xspace}

\newcommand{\weg}{\omega_{eg}}
\newcommand{\keg}{k_{eg}}

\newcommand{\Hc}{\mathrm{H.c.}}

\newcommand{\gN}{g\sqrt{N}}
\newcommand{\E}{\mathcal{E}}

\begin{document}
	%\printinunitsof{in}\prntlen{\linewidth}
	\title{Retrieval of single photons from solid-state quantum transducers}
	\author{Tom Schmit}
	\affiliation{Theoretical Physics, Department of Physics, Saarland University, 66123 Saarbr\"ucken, Germany}
	\author{Luigi Giannelli}
	\affiliation{Theoretical Physics, Department of Physics, Saarland University, 66123 Saarbr\"ucken, Germany}	
	\affiliation{Dipartimento di Fisica e Astronomia "Ettore Majorana", Universit\`a di Catania, Via S. Sofia 64, 95123 Catania, Italy}
	\affiliation{INFN, Sez. Catania, 95123 Catania, Italy}	
	\author{Anders S. S{\o}rensen}
        \affiliation{Center for Hybrid Quantum Networks (Hy-Q), Niels Bohr Institute,University of Copenhagen, Blegdamsvej 17, DK-2100 Copenhagen {\O}, Denmark}	
	\author{Giovanna Morigi}
	\affiliation{Theoretical Physics, Department of Physics, Saarland University, 66123 Saarbr\"ucken, Germany}

% - Abstract
\begin{abstract}
	Quantum networks using photonic channels require control of the interactions between the photons, carrying the information, and the elements comprising the nodes. In this work we theoretically analyse the spectral properties of an optical photon emitted by a solid-state quantum memory, which acts as a converter of a photon absorbed in another frequency range. We determine explicitly the expression connecting the stored and retrieved excitation taking into account possible mode and phase mismatch of the experimental setup. The expression we obtain describes the output field as a function of the input field for a transducer working over a wide range of frequencies, from optical-to-optical to microwave-to-optical. We apply this result to analyse the photon spectrum and the retrieval probability as a function of the optical depth for microwave-to-optical transduction. In the absence of losses, the efficiency of the solid-state quantum transducer is intrinsically determined by the capability of designing the retrieval process as the time-reversal of the storage dynamics.
\end{abstract}
\date{\today}

% - Title
\maketitle

\section{Introduction}
\label{Sec:Introduction}
Control of light-matter interactions is at the core of quantum technological applications~\cite{Acin:2018}. Its realization requires detailed understanding of photon absorption and emission processes in their microscopic details. This knowledge is a prerequisite for the implementation of quantum light sources for quantum sensors \cite{Gatti:2002,Cappellaro:2017} and for the realisation of protocols for quantum computation \cite{KLM:2001,Monroe:2010} and quantum communication \cite{Cirac:1997,Kimble:2008,Wehner:2018,Quantum-Crypto-Review:2002,Sangouard:2012,Brendel:1999}. Furthermore, control of the photon shape and frequency is essential for hybrid quantum networks, combining elements which work optimally in different frequency ranges~\cite{Cirac:1997,PNAS:2013,Stobinska:2009,SB:2018}.

\begin{figure}[!ht]
		\centering
		\includegraphics[width=\linewidth]{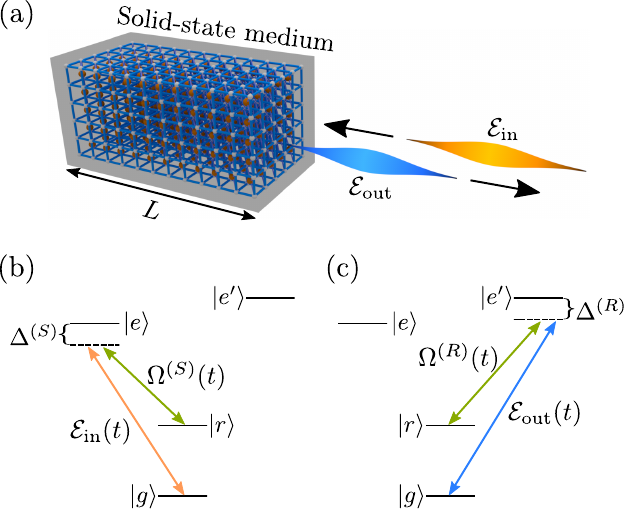}
		\caption{(color online) Single-photon frequency conversion with a solid-state quantum memory. Wave-mixing at the single-photon level is enabled by the emitters' internal level structure. (a) The medium consists of quantum emitters which interact with quantum fields in different frequency ranges. $L$ denotes the medium's length along the direction of photon propagation.  (b) The incident photon field $\E_{\rm in}$ couples to the transition $|g\rangle\to|e\rangle$ and the storage pulse $\Omega^{(S)}$ transfers the excitation into the stable state $|r\rangle$. (c) The excitation is then transferred by the read pulse $\Omega^{(R)}$ to a different state $|e'\rangle$, which performs a transition to the state $|g\rangle$ by emitting the photon field $\E_{\rm out}$. The frequency detunings $\Delta^{(S)}$ and $\Delta^{(R)}$ of the two transitions can be correlated. Write and read pulses can propagate at different group velocities than the incident and emitted photon, respectively.    
		}
		\label{fig:System_sketch}
	\end{figure}

In quantum networks, information is stored in stable quantum mechanical excitations, which constitute the quantum memory \cite{Afzelius2015}. In quantum memories for photons, for instance, the excitations can be the electronic transition of a single emitter \cite{Reiserer:2015,Duan:2010,SB:2018,Doherty:2013} or the spin wave of an ensemble of emitters \cite{Doherty:2013,Thiel:2011,Astner:2017,Distante:2017,Kuzmich:2016}. Among several realisations, solid-state quantum memories such as ensembles of Nitrogen-Vacancies in bulk diamond \cite{Doherty:2013} and rare-earth ion-doped crystals \cite{Thiel:2011} naturally provide large scattering cross sections and stable transitions~\cite{Lodahl:2019,CRIB-AFC-Review,Lauritzen:2012}. Rare-earth ion-doped crystals, moreover, can have level structures that allow one to implement quantum transducers for single photons, enabling the transfer of information between different frequency ranges by means of appropriate storage and retrieval protocols \cite{OBrien:2014,Blum:2015,Nadia:1,Lauk:2019,Li-Cappellaro:2019} acting on different transitions. For storage and retrieval on the same transition, the optimal memory efficiency is typically achieved by having the retrieval process being the time reverse of the storage dynamics \cite{Gorshkov:2007PRL}. %One limitation to the efficiency of existing protocols is the intrinsic inhomogeneous broadening of the emitters' frequencies \cite{CRIB-AFC-Review,Iakoupov:2013,Kroell:2001,Afzelius:2009,Gorshkov:2007PRL,Nadia:2}. 
Photon storage followed by retrieval on a different internal transition however, generally implies that the retrieval process cannot be cast in terms of the time reversed storage dynamics. Moreover, the level structure might not warrant the optimal conditions for realising the individual write or read protocols \cite{CRIB-AFC-Review,Iakoupov:2013,Kroell:2001,Afzelius:2009,Gorshkov:2007PRL}, such as for instance that incident photon and control pulses might have different wave numbers and group velocities.

In this work we theoretically analyse the dynamics of retrieval of single photons emitted by a solid-state quantum memory as a function of the stored excitation and of the memory properties. In particular, we focus on questions relevant for quantum transducers employing different transitions in the emitters. The setup we consider is illustrated in Fig. \ref{fig:System_sketch}. As opposed to previous studies, our analysis furthermore includes the effect of different group velocities and wave numbers between the photon and the read/write pulses. We finally analyse the efficiency of a solid-state quantum memory acting as a transducer from microwave to optical frequencies as in the protocols of Refs. \cite{OBrien:2014,Blum:2015}.

This manuscript is organized as follows. In Sec. \ref{Sec:Theory} we review the basic equations describing the dynamics of a single photon which propagates in a solid-state quantum memory and determine the general solution. In Sec. \ref{Sec:SR} we derive the equations which relate the spectrum of the emitted photon with the spectrum of the input photon for a generic distribution of the emitters within the bulk material. In Sec. \ref{Sec:The Protocol} we analyse the spectrum of a retrieved optical photon when the stored excitation is in the microwave regime. The conclusions are drawn in Sec. \ref{Sec:Conclusion and Outlook}. The appendices provide details of the model of Sec. \ref{Sec:Theory} and further details of the calculations presented in Sec. \ref{Sec:SR}.

\section{Single photon propagating in a solid-state medium}
	\label{Sec:Theory}
	
 In this section we review the basic equations describing the interactions of the quantum field with an ensemble of emitters and then derive their generic solutions. In our study we start from the models and equations of Ref. \cite{Gorshkov:2007PRL,Gorshkov:2007}. We note that in those treatments the dynamics of storage and retrieval are studied in a frame moving respectively with the write and read laser pulse. Here, we instead determine the expressions relating input field, stored excitation, and output field in the reference frame of the bulk embedding the emitters. This allows us to describe situations where different fields propagate at different group velocities.

\subsection{Equations for field and polarization}

We consider a solid-state quantum memory consisting of $N$ quantum emitters embedded in a bulk. The relevant internal levels of the emitters form a four-level system as in Fig. \ref{fig:System_sketch}. The ground state $\ket{g}$ couples to the states $|e\rangle$ and $|e'\rangle$ at transition frequencies $\omega_0^{(S)}$ and $\omega_0^{(R)}$, respectively. The transitions $|g\rangle\to |e\rangle$ and $|g\rangle\to |e'\rangle$ interact with photons propagating along the $z$-direction, while photon emission along other directions in space is included in the decay rate $\gamma$ ($\gamma'$) of the state $|e\rangle$ ($|e'\rangle$). 
%From now on we assume that the input photon couple with the transition $|g\rangle\to |e\rangle$. The absorbed photon is stored in the stable state  $|r\rangle$ by means of a pulse driving the transition $|e\rangle\to|r\rangle$. We denote this dynamics by storage and use the label $(S)$ in order to indicate its characteristic physical quantities. The retrieval dynamics ($(R$) consists of a second pulse transferring the excitation $|r\rangle$ to the state $|e'\rangle$, which in turn undergoes a transition to $|g\rangle$ by emitting the output photon. 

Below we introduce the basic equations describing the interaction between the transitions $|g\rangle\to |e\rangle$ and $|g\rangle\to |e'\rangle$ with the quantum fields. For $N\gg 1$ we introduce the operators $\op{}{P}{}^{(S)}(z,t;\Delta)$ and $\op{}{P}{}^{(R)}(z,t;\Delta)$ which describe the dynamics of an excitation in $|e\rangle$ and $|e'\rangle$, respectively. These operators give the polarization of the emitters at time $t$, detuning $\Delta$, and at the position $z$ on a coarse-grained spatial grid. The equations of motion are derived assuming small density fluctuations along $z$ over a fraction of the resonant wavelength, as we review in appendix \ref{App:A}. The polarization $\op{}{P}{}^{(j)}(z,t;\Delta)$ couples to the operators $\op{}{\E}{^{(j)}_f}(z,t)$ and $\op{}{\E}{^{(j)}_b}(z,t)$ describing the forward and backward propagating fields, respectively. Their dynamics is given by coupled Heisenberg equations of motion, which are linear in the operators. Since we will consider normally-ordered products, it is sufficient to consider the equations of motions of the corresponding complex amplitudes within the medium \cite{Gorshkov:2007PRL,Gorshkov:2007}. These are
 \begin{subequations} %  Checked
    	\label{alleqs:eoms_cf}
    	\begin{eqnarray}
        \left(\frac{\partial}{\partial t} + c^{(j)} \frac{\partial}{\partial z}\right)\E^{(j)}_f &=& i\mu_0^{(j)} L {\rm e}^{-{\rm i}k^{(j)} z}\\
        &\times& \int_{-\infty}^{+\infty}{\rm d}\Delta G^{(j)}(z,\Delta){P}{}^{(j)}(z,t;\Delta)\,,\nonumber
        	\label{eq:eom_E_cf}\\
        	\left(\frac{\partial}{\partial t} - c^{(j)} \frac{\partial}{\partial z}\right)\E^{(j)}_b &=& i\mu_0^{(j)} L {\rm e}^{{\rm i}k^{(j)} z}\label{eq:eom_E_cb}\\
        	&\times&\int_{-\infty}^{+\infty}{\rm d}\Delta G^{(j)}(z,\Delta){P}{}^{(j)}(z,t;\Delta)\,,\nonumber\\
        		\frac{\partial}{\partial t}{P}{}^{(S)} &=& -(\gamma/2  + {\rm i} \Delta) {P}{}^{(S)}+{\rm i}\mu_0^{(S)}\\
        		&&\times\left(\E^{(S)}_f{\rm e}^{{\rm i}k^{(S)}z}+\E^{(S)}_b{\rm e}^{-{\rm i}k^{(S)} z}\right)\,,\nonumber\\
        	\frac{\partial}{\partial t}{P}{}^{(R)} &=& -(\gamma'/2  + {\rm i} \Delta) {P}{}^{(R)}+{\rm i}\mu_0^{(R)}\\
        	&&\times\left(\E^{(R)}_f{\rm e}^{{\rm i}k^{(R)}z}+\E^{(R)}_b{\rm e}^{-{\rm i}k^{(R)} z}\right)\nonumber\,,
          	\label{eq:eom_P_cf}
    	\end{eqnarray}
    \end{subequations}
with $0 < z < L$, see appendix \ref{App:A} for details. Here, $c^{(j)}$ is the group velocity, $k^{(j)}>0$ is the wave number, and $\mu_0^{(j)}$ is the spectral bandwidth of the $j$-transition of the quantum memory ($j=S,R$)\cite{Iakoupov:2013}:
$$\mu_0^{(j)}=g^{(j)}\sqrt{N}\,,$$
where  $g^{(j)}$ is the (real-valued) coupling constant between a single emitter and the resonant field modes and the factor $\sqrt{N}$ accounts for the collective scattering. In the equations we have introduced the emitters' distribution $G^{(j)}(z,\Delta)$, which is a function of their position along the $z$-axis and of their frequency shift $\Delta$ from the average frequency $\omega_0^{(j)}$. This function may include both reversible and irreversible inhomogeneous broadening and is discussed in the following section. 

When analysing storage (retrieval) we will also consider a laser pulse coupling the state $|e\rangle$ ($|e'\rangle$) with a third stable level $|r\rangle$ as in the protocols of Refs. \cite{Kroell:2001}. We will assume that this control field implements a $\pi$-pulse, transferring the population from $|e\rangle$ to $|r\rangle$ or from $|r\rangle$ to $|e'\rangle$, on a time scale in which the coupling with the photon field can be neglected. Therefore, we can focus our analysis on Eqs. \eqref{alleqs:eoms_cf} and discuss separately the dynamics induced by the control field, see also Ref. \cite{Gorshkov:2007}. 

\subsection{Emitters' frequency distributions}

The distributions $G^{(j)}(z,\Delta)$ are normalized according to the relation
\begin{equation} %  Checked
	\label{eq:normalization_cond_G}
	\int_0^L{\rm d}z\int_{-\infty}^{+\infty}{\rm d}\Delta\, G^{(j)}(z,\Delta) = 1\,,
\end{equation}
with $z\in [0,L]$. The generalized dependence of $G^{(j)}(z,\Delta)$ on both variables $\Delta$ and $z$ allows one to describe all limiting cases considered in the literature, such as the Atomic Frequency Comb \cite{Afzelius:2009}, transverse CRIB \cite{Kroell:2001}, and longitudinal CRIB \cite{Moiseev:2008}.

For later convenience, we introduce the linear density of emitters $\tilde{n}(z)$, which is obtained by integrating the distribution $G$ over the frequencies:
\begin{equation} %  Checked
	\label{eq:ntilde:z}
	\tilde n^{(j)}(z)=\int_{-\infty}^{+\infty}{\rm d}\Delta\, G^{(j)}(z,\Delta)\,.
\end{equation}
The density of emitters at frequency $\Delta$ is instead defined as
\begin{equation} %  Checked
	\label{eq:n:Delta}
	 n^{(j)}(\Delta)=\int_0^L{\rm d}z \,G^{(j)}(z,\Delta)\,.
\end{equation}
Both densities are normalized to unity. A further useful quantity is the maximal value of the density $n^{(j)}(\Delta)$, which we denote by $n_0^{(j)}$:
\begin{equation} %  Checked
	n_0^{(j)}=\max_\Delta n^{(j)}(\Delta)\,.
\end{equation}
This quantity does not necessarily coincide with the density $n^{(j)}(0)$ at the central frequency. 

In general, one distinguishes reversible from intrinsic inhomogeneous broadening, where reversible inhomogeneous broadening is introduced and later reversed to control the emission process \cite{Kroell:2001,Afzelius:2009,Gorshkov:2007PRL,Moiseev:2008} while intrinsic inhomogeneous broadening is a medium's property which limits the performance of quantum memory protocols. These different types of broadening can be included in this formalism by writing $\Delta=\Delta_0+\Delta_1$ where $\Delta_\ell$ follows the distribution $G_\ell^{(j)}(z,\Delta_\ell)$ and $\ell=0$ ($\ell=1$) labels the reversible (irreversible) broadening. Correspondingly,
\begin{eqnarray} %  Checked
	G^{(j)}(z,\Delta)&=&\int_{-\infty}^{+\infty}{\rm d}\Delta_0\int_{-\infty}^{+\infty}{\rm d}\Delta_1 \label{irreversible}\\
	 & & \times G_0^{(j)}(z,\Delta_0) G_1^{(j)}(z,\Delta_1)\delta(\Delta-\Delta_0-\Delta_1)\nonumber\,.
\end{eqnarray} 
In several works the  overall effect of the intrinsic inhomogeneous broadening is phenomenologically described by an effective dephasing rate \cite{CRIB-AFC-Review,OBrien:2014}. In the present paper the generic distribution $G^{(j)}(z,\Delta)$ includes both contributions, unless specified otherwise.

\subsection{Formal solution}

We analyse the solutions of Eqs. \eqref{alleqs:eoms_cf} by taking the Laplace transform with respect to time. In the following we omit the superscript $(j)$ for brevity and write $\gamma = \gamma'$. Let $\bar{X}(u)$ be the Laplace transform of the function $X(t)$, such that $\bar{X}(u)=\int_{t_0}^\infty {\rm d}t\exp(-ut)X(t)$. Here, $t_0$ is an initial time. The Laplace transform of the polarization is given by the equation
\begin{equation} %  Checked  
	\label{bar:P:u}
	\bar P(z,u;\Delta)=\frac{P(z,t_0;\Delta){\mathrm{e}^{-ut_0}}}{u+{\rm i}\Delta+\gamma/2}+{\rm i}\mu_0\,\frac{\bar{\E}_f {\rm e}^{{\rm i}k z} + \bar{\E}_b{\rm e}^{-{\rm i}k z}}{u+{\rm i}\Delta+\gamma/2}\,,
\end{equation}  
and is a function of the polarization $P(z,t_0;\Delta)$ at $t=t_0$ and of the fields within the medium. 
The Laplace transforms of the backward and forward propagating fields read
\begin{eqnarray} %  Checked    
	\bar{\E}_f(z,u) &=& \bar{\E}_f(0,u){\rm e}^{-H_f(z,u)}+\int_{0}^z\frac{\mathrm{d}z'}{c}\mathrm{e}^{-(H_f(z,u)-H_f(z',u))}\nonumber\\
	&&\times\left\{\E_f(z',t_0)+{\rm i}\mu_0L\mathcal{P}_G(z',u){\rm e}^{-{\rm i}k z'}\right\}{\rm e}^{-ut_0}\,,\nonumber\\
	\label{E:f:u}\\
	\bar{\E}_b(z,u) &=& \bar{\E}_b(L,u){\rm e}^{-H_b(z,u)}+\int_{z}^L\frac{\mathrm{d}z'}{c}\mathrm{e}^{-(H_b(z,u)-H_b(z',u))}\nonumber\\
	&&\times\left\{\E_b(z',t_0)+{\rm i}\mu_0L\mathcal{P}_G(z',u){\rm e}^{{\rm i}k z'}\right\}\mathrm{e}^{-ut_0}\,,\nonumber\\
	\label{E:b:u}
\end{eqnarray}
where $ \bar{\E}_f(0,u)$ and $ \bar{\E}_b(L,u)$ are the Laplace components at the medium edges and
\begin{eqnarray} %  Checked  
	\mathcal P_G(z,u)&=& \int_{-\infty}^{+\infty}{\rm d}\Delta\frac{G(z,\Delta)}{u+{\rm i}\Delta+\gamma/2}P(z,{t_0};\Delta)\,.
\end{eqnarray}
Here, we have discarded the coupling between the fields $\E_f$ and $\E_b$ mediated by the polarization. This approximation requires that the coherence length of the incident photon inside the medium is much longer than the photon's wavelength so that the emitters' distribution varies slowly on the scale of a wavelength.
\begin{eqnarray} %  Checked    
	\label{H:b}
	H_b(z,u)&=&\frac{u(L-z)}{c}+d\, h(z,u)\,,
\end{eqnarray}
with
\begin{equation} %  Checked  
	\label{Eq:h}
	h(z,u)=\frac{1}{2\pi n_0}\int_{z}^{L}{\rm d}z'\int_{-\infty}^{+\infty}{\rm d}\Delta\frac{G(z',\Delta)}{u+{\rm i}\Delta+\gamma/2}\,,
\end{equation}
and $H_f(z,u)=H_b(0,u)-H_b(z,u)$. These functions depend on the dimensionless parameter 
\begin{equation}  %  Checked    
\label{d:prime}
	d=\frac{2\pi \mu_0^2 n_0}{c}L\,,
\end{equation}
which we denote here by "optical depth". The parameter $d$ determines the attenuation of an incoming field that propagates through the medium, see Sec. \ref{sec:transmitted_field}. This definition is convenient when the broadening is larger than the damping $\gamma$, as we assume in this work, and it matches the standard definition of the experimentally observed optical depth (i.e. $d^{(S)}$ corresponds to $2d'$ in Ref. \cite{Gorshkov:2007IB}). For a narrow distribution, however, the expression in Eq. \eqref{d:prime} does not correspond to the observed optical depth.
\section{Storage and retrieval of a single photon}
\label{Sec:SR}

In this section we analyse the dynamics and efficiency of photon storage and retrieval using Eqs. \eqref{bar:P:u}-\eqref{E:b:u}. Storage and retrieval is here implemented by means of the fast protocol of Refs. \cite{Kroell:2001,Gorshkov:2007}, 
where population is transferred between the stable ground state $\ket{r}$ and the excited states $|e\rangle$ and $|e'\rangle$ by means of fast resonant pulses.
%where population is transferred from the state $|e\rangle$ to a stable state $|r\rangle$ by a very fast resonant pulse. 
 
\subsection{Storage}

We consider a photon wave packet which propagates along the negative direction of the $z$-axis. In the following, we denote the emitters' distribution by $G^{(S)}(z,\Delta)$ and use the superscript $(S)$ to indicate the storage dynamics. In this model the input photon is described by a complex field $\E_{\rm in}(t)$ at the position $z=L$ of the medium and propagating in the backward direction:
$$
	\E_b^{(S)}(L,t)=\E_{\rm in}(t)\,,
$$
for $t\ge t_0$ while at $t=t_0$ fields and polarization vanish inside the medium. The Laplace component of the backward field is given by
\begin{eqnarray}  
	\label{bar:E:s}
	\bar\E_b^{(S)}(z,u) &=&{\bar{\E}_\mathrm{in}(u)}{\rm e}^{-H_b^{(S)}(z,u)}\,,
\end{eqnarray}
where $\bar{\E}_\mathrm{in}(u)$ is the Laplace transform of the input field. The Laplace component of the polarization then reads
\begin{equation}  
	\label{bar:P}
	\bar P^{(S)}(z,u;\Delta)={\rm i}\mu_0^{(S)}\,\frac{\bar{\E}_b^{(S)}(z,u){e}^{-ik^{(S)} z}}{u+{\rm i}\Delta+\gamma/2}\,.
\end{equation}  

The polarization at the instant of time $t_S>t_0$ is the inverse Laplace transform of Eq. \eqref{bar:P}. We determine it using Eq. \eqref{bar:E:s} in Eq. \eqref{bar:P} and making the reasonable assumption that the distribution $G^{(S)}(z,\Delta)$ identically vanishes for $|\Delta|>\Delta_{\rm max}>0$, \ie, $G^{(S)}(z,\Delta)$ is different from zero only in the finite frequency interval $[-\Delta_{\rm max}, \Delta_{\rm max}]$ for some maximum detuning $\Delta_{\rm max}$. This allows us to perform the inverse Laplace transform of Eq. \eqref{bar:P}:%We take the inverse Laplace transform of Eqs. \eqref{bar:P} and \eqref{bar:E:s} \blue{to} determine the polarization and the backward field inside the medium at the instant of time $t=t_S$.
\begin{eqnarray}  
	\label{P:S}
	P^{(S)}(z,t_S;\Delta) &=& \mathrm{e}^{-ik^{(S)}z}\int_{t_0}^{t_S}\mathrm{d}\tau \mathcal{F}^{(S)}(z,t_S-\tau;\Delta)\E_\mathrm{in}(\tau)\,.\nonumber\\
\end{eqnarray}
The polarization is thus the convolution integral of the input field with the function $\mathcal{F}^{(S)}$. The function $\mathcal{F}^{(S)}$, in turn, describes the response of the medium and is given by the integral 
\begin{eqnarray}  
	\mathcal{F}^{(S)}(z,t;\Delta) &=& \frac{\mu_0^{(S)}}{2\pi}\int_{C}{\rm d}u \frac{{\rm e}^{ut}\mathrm{e}^{-H^{(S)}_b(z,u)}}{u+{\rm i}\Delta+\gamma/2}\label{eq:F_S:z,u,Delta}\,,
\end{eqnarray}
where $C$ is the path in the complex plane along the Bromwich contour. We remark that $\mathcal{F}^{(S)}(z,t;\Delta)$ vanishes for $t<(L-z)/c^{(S)}$, consistent with causality arguments.

\subsubsection{Fast storage}

Perfect storage is achieved when the backward field is completely mapped onto the polarization. A consequence is that at a given instant of time the field inside the medium must vanish. Following the fast protocol of Ref. \cite{Gorshkov:2007}, at this instant of time a fast control pulse transfers the excitation to a third, metastable level $|r\rangle$. Let $S(z,t;\Delta)$ denote the corresponding coarse-grained spin wave, and $k'^{(S)}$ be the wave number of the transition $|e\rangle\to |r\rangle$ at the central frequency. Then the polarization in Eq. \eqref{P:S} is mapped into the spin wave if the pulse area is $\pi$ and the pulse duration is much smaller than the temporal width of the photon. The resulting spin wave reads (see appendix \ref{App:SpinWave} for details)
\begin{eqnarray}
\label{Spin:Storage:1}
S(z,t_0(z)^+;\Delta)=-{\rm i} {\rm e}^{{\rm i}k^{\prime(S)}z} P^{(S)}(z,t_0(z);\Delta)\,,\nonumber\\
\end{eqnarray}		
where $t_0(z)=t_0+\delta t+(L-z)/c'^{(S)}$, $\delta t$ is the time delay of the laser control pulse at $z=L$, and $c'^{(S)}$ is the group velocity in the corresponding frequency range. We note here that the classical control field may have, e.g.,  a different polarization  than the quantum field. This means that if light is guided by an asymmetric waveguide of sufficiently small size, the quantum and control fields can have substantially different velocities in the waveguide. In the above expression we have assumed that the transfer is perfect and instantaneous over the time scale of the photon dynamics. If this transfer is not optimal, some population will remain in the state $|e\rangle$ and will be lost by damping. 

We write the spin wave as a function of the incident field using Eq. \eqref{P:S} in Eq. \eqref{Spin:Storage:1}:
\begin{eqnarray}
\label{Spin:Storage}
&&S(z,t_0(z)^+;\Delta)=-{\rm i} \frac{\mu_0^{(S)}}{2\pi} {\rm e}^{{\rm i}\delta k^{(S)}z} \\
& \times & \int_{C}{\rm d}u \frac{{\rm e}^{u(t_0+\delta t)}\mathrm{e}^{u(L-z)/c_{\rm eff}^{(S)}}\mathrm{e}^{-d^{(S)}h^{(S)}(z,u)}}{u+{\rm i}\Delta+\gamma/2}\bar\E_{\rm in}(u)\nonumber\,,
\end{eqnarray}		
with $\delta k^{(j)}=k^{\prime(j)}-k^{(j)}$. The parameter $c_{\rm eff}^{(j)}$ is an effective group velocity, defined as
\begin{equation}
\frac{1}{c_{\rm eff}^{(j)}}=\frac{1}{c^{\prime(j)}}-\frac{1}{c^{(j)}}\,,
\end{equation}
that can take both positive and negative values. %The term $1/{c_{\rm eff}}$ vanishes when $c=c'$. 

\subsubsection{Transmitted field}
\label{sec:transmitted_field}
Perfect storage implies that the field is absorbed by the medium, and thus the intensity at the opposite edge of the medium must vanish. In order to determine the intensity $\mathcal I_0$ at $z=0$, we note that the Laplace component of the field corresponds to the Fourier component, namely $$\bar{\E}_b^{(S)}(z,u)|_{u=-i\omega}=\sqrt{2\pi}\,\tilde{\E}_b^{(S)}(z,\omega)\,,$$ where $\tilde{\E}_b^{(S)}(z,\omega)$ is the field Fourier component at position $z$. Therefore, the intensity $\mathcal I_0$ at $z=0$ takes the form
$$\mathcal I_0=\int{\rm d}\omega \mathcal I(\omega)\,,$$ where $\mathcal I(\omega)=|\tilde{\E}_b^{(S)}(0,\omega)|^2$ is the spectral component. Using Eq. \eqref{bar:E:s} we find
\begin{eqnarray}  
	\label{eq:Eout_storage:w}
	\mathcal I(\omega) &=& |\tilde{\E}_\mathrm{in}(\omega)|^2\\
	&&\times\exp\left(-d^{(S)}\int_{-\infty}^{+\infty}\frac{\mathrm{d}\Delta}{2\pi} \frac{\gamma\, n^{(S)}(\Delta)/n_0^{(S)}}{(\Delta-\omega)^2+(\gamma/2)^2}\right)\nonumber\,,
\end{eqnarray}
with $d^{(S)}$ being the optical depth of Eq. \eqref{d:prime}. The field vanishes at the medium edge $z=0$ when $\mathcal I(\omega) \approx 0$ for all frequency components. As seen from Eq. \eqref{eq:Eout_storage:w}, in a broadened medium this leads to the inequality $d^{(S)} n^{(S)}(\omega)/n_0^{(S)}\gg 1$ for the frequencies $\omega$ of the photon wave packet.
In essence this expresses that for a broadened medium an efficient memory can only be attained if there is a sufficiently large optical depth at the frequency of the incoming pulse.  (We note that this is different for a homogeneously broadened medium  where only the resonant optical depth needs to be large \cite{Gorshkov:2007PRL}). 
Furthermore, Eq. \eqref{eq:F_S:z,u,Delta} enables one to identify the spatial size of the region where the photon is stored in the medium. For a uniform distribution $G^{(S)}$, the size is of the order of $L/d^{(S)}$. For $d^{(S)}\gg 1$ the photon is hence stored within a relatively small region close to the edge where it has entered the medium. The fraction of radiation that is lost by decay is of the order of $\gamma T_\mathrm{coh}$, where  $T_\mathrm{coh}$  is the photon's coherence time. In this work we consider photons with relatively large spectral width $\delta\omega$, therefore $\gamma T_\mathrm{coh}\ll 1$.

%For this reason higher retrieval efficiencies are typically achieved by coupling the polarization with the field propagating into the opposite direction than the incident photon \cite{Kroell:2001,Afzelius:2009,Gorshkov:2007}.
%Finally, from the above argument we infer that $dc/L$ determines the characteristic time scale over which the incident photon is absorbed. Thus, damping processes can be discarded provided that the condition $\gamma L/c \ll d$ is satisfied. 

%Hence it we use short pulses the only limitation is what is lost. On the other hand we assume that d is constant, so this only applies to pulses more narrow than the width of the broadening.   

%Damping can be neglected assuming that the interaction time $T_\mathrm{int}$ fulfils $T_\mathrm{int}\gamma \ll 1$. We estimate the interaction time to be given by the photon's coherence time $T_\mathrm{coh}$ (here a pure state \cite{Mueller:2017}), $T_\mathrm{int} \sim T_\mathrm{coh}$, assuming that the coherence length of the incident photon in free space is larger than the medium length $L$ \cite{Afzelius:2009}. This leads to the condition $\gamma T_\mathrm{coh}\ll 1$.

%and is here introduced for convenience. In the presence of broadening and off-resonance excitation the optical depth of the ensemble is given by the relation (see Eq. \eqref{eq:Eout_storage:w}):
%$$d'(\omega)=d\int_{-\infty}^{+\infty}\frac{\mathrm{d}\Delta}{2\pi} \frac{\gamma\, n(\Delta)/n_0}{(\Delta-\omega)^2+(\gamma/2)^2}\,,$$
%where $d'<d$ and depends on the specific form of the distribution $G$. See also Ref. \cite{Gorshkov:2007IB}.

\subsection{\label{sec:retieval} Retrieval}

We now assume that at a given time after the photon has been stored another control pulse enters the medium in the forward propagating direction (and thus counterpropagating with respect to the direction of incidence of the initial photon). The pulse has ideally an area of $\pi$ for all emitters and transfers the spin-wave excitation into the polarization. In CRIB protocols, moreover, an effective Hahn echo is implemented on the emitters' frequency distribution, ideally performing the transformation $\Delta\to -\Delta$. The underlying assumption is that the inhomogeneous broadening of storage and retrieval transitions are correlated and reversible since they are induced by some external field. This may or may not be true for a transducer where the electronic transitions for storage and retrieval can be different. In this case the correlations  will depend on the microscopic mechanisms responsible for the broadening. 

In the following we denote the transformation of the broadening between storage and retrieval by the generic map $$\Delta\to p[\Delta]\,,$$ which possibly includes imperfections in the realisation. The absence of correlations between storage and retrieval is recovered for $p[\Delta]$ taking random values within a given distribution, see appendix \ref{App:different_broadenings}. 

The field emitted by the medium at $z=L$, $\bar{\E}_{\rm out}(u)=\bar{\E}_f^{(R)} (L,u)$, depends on the stored polarization:
\begin{eqnarray}
P^{(R)}(z,t_1(z)^+;{p[\Delta]})= - {\rm e}^{{\rm i}(k'^{(S)}+k'^{(R)})z}P^{(S)}(z,t_0(z);\Delta)\,.\nonumber\\
\label{P:R}
\end{eqnarray}
Here $t_1(z)=T_1+T_S+z/c'^{(R)}$, where $T_1$ is the instant of time at which the photon has been stored (see appendix \ref{App:SpinWave} for details) and $T_S$ is the storage time.% The counterpropagating pulse imprints a phase grating, such that for $k'^{(R)}+k'^{(S)}=k^{(R)}+k^{(S)}$ emission into the backward propagating field (which here coincides with the direction of the incident field) is suppressed. On the other hand, the output in the forward direction of the counterpropagating pulse (opposite to the original incident pulse) is enhanced for $|k'^{(R)}+k'^{(S)}|=|k^{(R)}+k^{(S)}|$ since then the contributions for different $z$, c.f. Eq. \eqref{E:f:u}, add up coherently.

%\begin{eqnarray}  
%	\bar\Eout(u) &=& \frac{L}c\int_{0}^L\mathrm{d}z{\rm e}^{-ik^{(R)}z}\int_{-\infty}^{+\infty}\mathrm{d}\Delta G^{(R)}(z,\Delta)\nonumber\\
%	& &\times P^{(R)}(t_1(z),z;\Delta)\,.
%	\label{E:out:P}
%\end{eqnarray}
%
Using Eq. \eqref{P:R} one can now determine the explicit relation between input and output photon. For this purpose, we first observe that, since the forward field vanishes before the retrieval, the Fourier component $\tilde{\E}_\mathrm{out}(\omega)$  corresponds  to the Laplace component $\bar\E_{\rm out}(u)$ taken at $u=-{\rm i}\omega$, apart from a normalization factor. We now use Eqs. \eqref{bar:E:s}, \eqref{bar:P}, and \eqref{P:R} in Eq. \eqref{E:f:u} and obtain the integral relation \cite{Nunn:2008}
\begin{eqnarray}  
	    \label{b:E:out}
		\tilde{\E}_{\rm out}(\omega)=\frac{1}{2\pi}\int_{-\infty}^{+\infty}\mathrm{d}\omega'\mathcal{S}(\omega,\omega')\tilde{\E}_{\rm in}(\omega')\,,
\end{eqnarray}
which connects the spectrum of the retrieved photon with the Fourier component of the input field $\tilde{\E}_\mathrm{in}(\omega)$. The kernel $\mathcal{S}(\omega,\omega')$ is a function of the emitters' distributions $G^{(R)}$ and $G^{(S)}$ and takes the explicit form
\begin{widetext} 
	\begin{eqnarray}
		\label{eq:F:v,u}
		\mathcal{S}(\omega,\omega') &=& \frac{\sqrt{d^{(S)}d^{(R)}}}{2\pi n_0}\sqrt{\frac{c^{(S)}}{c^{(R)}}} {\rm e}^{{\rm i}\omega(T_1+T_S+L/c'^{(R)})}{\rm e}^{-{\rm i}\omega'(T_1-L/c'^{(S)})}\\
		& &\times \int_{0}^{L}{\rm d}z\,{\rm e}^{{\rm i}(\delta k^{(S)}+\delta k^{(R)})z}\int_{-\infty}^{+\infty}{\rm d}\Delta \,\,\frac{G^{(R)}(z,p[\Delta])\,{\rm e}^{-{\rm i}\omega (L-z)/c_{\rm eff}^{(R)}}{\rm e}^{-d^{(R)} h^{(R)}(z,-{\rm i}\omega)}{\rm e}^{-{\rm i}\omega'(L-z)/c_{\rm eff}^{(S)}}{\rm e}^{-d^{(S)} h^{(S)}(z,-{\rm i}\omega')}}{({\rm i}(p[\Delta]-\omega)+\gamma'/2)({\rm i}(\Delta-\omega')+\gamma/2)}\,.\nonumber
	\end{eqnarray}
\end{widetext}
%where for simplicity we have assumed that $n_0$ is the same in the storage and retrieval dynamics. Expression \eqref{eq:F:v,u} accounts for the possible asymmetry in the dynamics  of storage and retrieval (including the irreversible component of the inhomogeneous broadening). It also accounds for the finite spectral width of the emitters' distribution.
We define the efficiency of the quantum transducer as the ratio between the number of outgoing and incoming photons \cite{Kroell:2001,Vivoli2013}:
\begin{equation}
	\label{eq:efficiency}
	\eta  = \frac{\int_{-\infty}^{+\infty}\mathrm{d}\omega|\tilde{\E}_\mathrm{out}(\omega)|^2}{\int_{-\infty}^{+\infty}\mathrm{d}\omega|\tilde{\E}_\mathrm{in}(\omega)|^2}\,.
\end{equation}
This definition ensures that the transducer has non-zero quantum capacitance as soon as the efficiency exceeds $\eta = 0.5$ \cite{Wolf:2007}, since our model accounts only for amplitude damping \cite{Gorshkov:2007}. We note that, even for efficiencies below the threshold 0.5, the transducer can still be used for non-trivial quantum information tasks by post selecting successful events \cite{Zeuthen:2020}.

Retrieval protocols for quantum memories achieve largest efficiency when the retrieval process is effectively the time-reversed storage dynamics \cite{Gorshkov:2007PRL}. The efficiency is naturally going to be reduced when the process of storage and retrieval are quite different, such as in the case of a transducer. In the case of a solid-state medium, the internal transitions coupling with the incident field are different from the ones coupling with the emitted field. Correspondingly, the spectral bandwidths of the quantum memories can be different. In Eq. \eqref{eq:F:v,u}, for instance, this could lead to different optical depths for the storage and retrieval process. Moreover, depending on the configuration, the group velocity of the reading laser pulse can substantially differ from the group velocity of the emitted photon. Finally, the wave number $k'^{(R)}$ of the reading pulse will be generally different from the wave number $k^{(R)}$ of the photon field. All these effects are accounted for in Eq. \eqref{eq:F:v,u}. We remark that in Eq. \eqref{eq:F:v,u} we have taken the same maximal value of the density $n_0$ for storage and retrieval. Nevertheless, possible differences only amount to a rescaling of the distribution $G^{(R)}$. From now on we will also assume that $\gamma'=\gamma$. Differences in the damping mean that the losses from a finite pulse duration are different. Here we are mainly interested in the limitations from mode matching and we therefore restrict ourselves to this simplified description.

\subsubsection*{Spatially-independent inhomogeneous broadening}

 In order to provide an example, in what follows we determine  the form of Eq. \eqref{eq:F:v,u} for the case of the CRIB protocol when the emitters' distribution does not depend on the position $z$ along the medium, $G^{(j)}(\Delta)=n^{(j)}(\Delta)/L$, but the distribution for storage and retrieval are correlated \cite{Afzelius:2009,Gorshkov:2007IB}. Now Eq. \eqref{Eq:h} can be cast in the convenient form 
\begin{equation}
h(z,-{\rm i}\omega)=\mathcal H(\omega)\,\frac{L-z}L\approx \frac{L-z}{2L}\,,
\end{equation}
where 
\begin{equation}  
	\label{H:omega}
	\mathcal{H}(\omega) = \frac{1}{2\pi n_0}\int_{-\infty}^{+\infty}\mathrm{d}\Delta\frac{n(\Delta)}{{\rm i}(\Delta-\omega)+\gamma/2}\,.
\end{equation}
This function takes the value $1/2$ when the emitter density is constant over the spectral width of the photon \cite{Kroell:2001,Afzelius:2009}, which is the case we consider here.

We now consider the map $p[\Delta]=-\Delta$, assuming that the inhomogeneous broadening is perfectly reversible. After performing the integrals in Eq. \eqref{eq:F:v,u}, the kernel becomes
\begin{widetext} 
	\begin{eqnarray}
		\label{eq:kernel:t}
		\mathcal{S}(\omega,\omega') &=& {\rm e}^{{\rm i}\omega(T_S+L/c'^{(R)})}{\rm e}^{{\rm i}\omega'L/c'^{(S)}} {\rm e}^{{\rm i}(\delta k^{(R)}+\delta k^{(S)})L}\,\frac{\sqrt{d^{(S)}d^{(R)}c^{(S)}/c^{(R)}}}{\gamma-{\rm i}(\omega+\omega')}\,\cdot\,\frac{1-{\rm e}^{-{\rm i}\mathcal F(\omega,\omega')}{\rm e}^{-(d^{(R)}+d^{(S)})/2}}{(d^{(R)}+d^{(S)})/2+{\rm i}\mathcal F(\omega,\omega')}\,,
		%{\rm e}^{-{\rm i}\omega (L-z)/c_{\rm eff}}e^{-d h_R(z,-{\rm i}\omega)}{\rm e}^{-{\rm i}\omega'(L-z)/c_{\rm eff}}{\rm e}^{-d h_S(z,-{\rm i}\omega')}\,,\nonumber
	\end{eqnarray}
\end{widetext}
where we have chosen the initial time $t_0$ such that $T_1=0$ and we have introduced the cutoff function
\begin{equation}
\label{eq:F(w,w')}
\mathcal F(\omega,\omega')=(\delta k^{(R)}+\delta k^{(S)})L+2\pi(\omega/\omega^{(R)}+\omega'/\omega^{(S)})
\end{equation}
and the characteristic frequency ($j=R,S$): $$\omega^{(j)}= 2\pi c_{{\rm eff}}^{(j)}/L\,.$$ 

In order to understand the effect of the individual components in Eq. \eqref{eq:kernel:t} let us first consider retrieval of the stored photon when $|e\rangle=|e'\rangle$. We assume mode matching, $\delta k=0$, and non-vanishing characteristic frequencies, $\omega^{(S)}=\omega^{(R)}=\omega^{(0)}$, which corresponds to light propagation in an asymmetric waveguide of sufficiently small size. In this case
	\begin{eqnarray}
		\label{eq:kernel:t_Lambda}
		\mathcal{S}_0(\omega,\omega') &=& {\rm e}^{{\rm i}\omega T_S}{\rm e}^{{\rm i}(\omega'+\omega)L/c'}\,\frac{1}{\gamma-{\rm i}(\omega+\omega')}\,\\
		& &\times\,\frac{1-{\rm e}^{-{\rm i}2\pi(\omega+\omega')/\omega^{(0)}}{\rm e}^{-d}}{1+{\rm i}2\pi(\omega+\omega')/(d\omega^{(0)})}\,.\nonumber
		%{\rm e}^{-{\rm i}\omega (L-z)/c_{\rm eff}}e^{-d h_R(z,-{\rm i}\omega)}{\rm e}^{-{\rm i}\omega'(L-z)/c_{\rm eff}}{\rm e}^{-d h_S(z,-{\rm i}\omega')}\,,\nonumber
	\end{eqnarray}
For $2\pi\gamma/(d\omega^{(0)})\ll 1$ the term $(\gamma-{\rm i}(\omega+\omega'))^{-1}$ behaves like a Dirac-delta function and one recovers the ideal case: the spectrum of the emitted photon is equal to the spectrum of the stored one and the single-photon retrieval probability, 
\begin{equation}
	\label{eq:Pemission}
	\mathcal W  = \int_{-\infty}^{+\infty}\mathrm{d}\omega|\tilde{\E}_\mathrm{out}(\omega)|^2 \,,
\end{equation}
approaches unity for $d\to\infty$ \cite{Kroell:2001}. This quantity coincides with the efficiency $\eta$, Eq. \eqref{eq:efficiency}, when the incident field is a single photon, namely, when $\int_{-\infty}^{+\infty}\mathrm{d}\omega|\tilde{\E}_\mathrm{in}(\omega)|^2=1$.

Mode mismatching, $\delta k\neq 0$, determines a characteristic length that shall be compared with the size of the spatial region where the photon is stored, $L/d$. When $|\delta k| L/d\ll 1$ this effect can be discarded. Rare-earth ion-doped crystals employed for realizing quantum memories typically have lengths $L$ of a few millimeters and transition frequencies between the hyperfine ground states representing $\ket{g}$ and $\ket{r}$ of tens of MHz to a few GHz, cf. for instance Refs. \cite{Nadia:1,Nadia:2,Afzelius:Eu}. Assuming a phase velocity of the order of the speed of light, $|\delta k| L$ takes values ranging from $10^{-2}$ to $10^{-1}$ and the regime $|\delta k| L/d \ll 1$ is achieved for optical depths of the order of unity. This regime is experimentally realised, for instance in the setups of Refs. \cite{Thiel:2011,Lauritzen:2012,Afzelius:Eu,Nadia:1,Nadia:2}.

Deviations from these optimal conditions are found when the storage and retrieval processes are characterized by different parameters, as is the case for a quantum transducer. We first note that the phase-matching condition now requires minimizing the quantity $\delta k^{(R)}+\delta k^{(S)}$, which, depending on the specific scheme, might be even achieved by retrieving the photon in the same direction of incidence. We note, however, that if the storage and retrieval are in the same direction, the pulse will have to travel through an optically dense medium which reduces the possible efficiency \cite{Afzelius:2009}. Alternatively, phase matching could also be achieved by suitably choosing the propagation direction of the write and read pulses \cite{Damon:2011}.

In order to highlight the role of asymmetry we assume phase matching $\delta k^{(R)}+\delta k^{(S)}=0$, $\omega^{(R)}=\omega^{(S)}=\omega^{(0)}$, and $c^{(S)}=c^{(R)}$, but different optical depths, $d^{(R)} \neq d^{(S)}$. We define the average optical depth $\bar d=(d^{(R)}+d^{(S)})/2$ and observe that the kernel can now be written in the same form as the ideal case, Eq. \eqref{eq:kernel:t_Lambda}, with $d\to\bar d$, but multiplied by the overall factor $\sqrt{d^{(R)}d^{(S)}}/\bar d$. This factor leads to the scaling of the retrieval probability
$$
	\mathcal W\propto d^{(R)}d^{(S)}/\bar{d}^2\,.
$$
The retrieval probability is thus reduced in schemes with very different optical depths for storage and retrieval. 
The inequality $\omega^{(R)}\neq\omega^{(S)}$ leads to a further reduction of the retrieval probability and modifies the spectrum of the output photon. The impact of this term scales with $(\omega^{(R)}-\omega^{(S)})/(\bar d\omega^{(R)}\omega^{(S)})$ and is reduced at sufficiently large average optical densities.  

In summary, the quantum transducer here discussed is realised by combining the dynamics of two quantum memories working in two different frequency ranges. As such, there are two kinds of frequency cutoffs:
the cutoffs to the emitters' frequency distributions and the frequency cutoffs $\omega^{(S)},\omega^{(R)}$ entering the function $\mathcal{F}(\omega,\omega^\prime)$, Eq. \eqref{eq:F(w,w')}. Both limit the efficiency and affect the spectrum of the retrieved photon. The photon spectral width shall thus be smaller than these cutoffs, whose minimum is the conversion bandwidth of the quantum transducer. Moreover, if the bandwidth of the storage and retrieval processes are different, the transducer can also act as a bandwidth transducer, converting the bandwidth of the photon from the storage to the (smaller) retrieval bandwidth.
\section{\label{Sec:The Protocol}Microwave-to-optical conversion}

In this section we analyse the retrieval dynamics when the storage process is done directly on the microwave transition between the stable states $|g\rangle$ and $|r\rangle$ in Fig. \ref{fig:System_sketch}. Several protocols for quantum memories in the microwave regime use the coupling with a single-mode resonator \cite{OBrien:2014,Blum:2015,Li-Cappellaro:2019}, where the characteristic wavelengths are larger than the size of the medium. Under these conditions we assume that the microwave excitation is uniformly distributed along the medium and that a fast $\pi$-pulse propagating in the forward direction transfers it to the optical polarization, giving
\begin{eqnarray}
\label{P:R:mw}
P^{(R)}(z,t_1(z)^+;\Delta)&=&{\rm e}^{{\rm i}k'z}f(\Delta)\,.
\end{eqnarray}		
Here $t_1(z)=z/c'^{(R)}$ and $f(\Delta)$ is a complex function that solely depends on $\Delta$. It fulfils the normalization condition
\begin{equation}
\label{norm:f}
\frac{L}{c^{(R)}}\int_{0}^{L}\mathrm{d}z\int_{-\infty}^{+\infty}\mathrm{d}\Delta G^{(R)}(z,\Delta)|f(\Delta)|^2 = 1\,,
\end{equation}
corresponding to a perfect transfer from the spin wave. The wave number $k'$ here also includes any possible contribution to the phase grating by the microwave storage process. We assume that the function $f(\Delta)$ can depend on the inhomogeneous broadening of the optical transition. This can for instance be the case if there are correlations between the emitters' frequency distributions during the microwave storage and optical retrieval. Alternatively a non-trivial function $f(\Delta)$ may also arise if the broadening plays a role during the optical $\pi$-pulse. For a $\pi$-pulse of non-vanishing duration the transfer to the excited state will be influenced by the broadening, e.g., excitations stored in state $|r\rangle$ cannot be transferred to the excited state $|e'\rangle$ if it is too far detuned. This will lead both to a detuning dependent amplitude $f(\Delta)$ of the excitation and to a reduction of the transfer efficiency. The latter can be described by including a rescaling factor smaller than unity in Eq. \eqref{norm:f}. A full investigation of the excitation dynamics is beyond the scope of this work. For now we therefore restrict ourselves to the effect of the shape of $f(\Delta)$. Note that any reduction in efficiency due to finite excitation efficiency can always be accounted for by multiplying our results by that efficiency.
 
The Fourier component at frequency $\omega$ of the output field is found by using Eq. \eqref{P:R:mw} in Eq. \eqref{E:f:u} as well as the correspondence with the Laplace component. It takes the form
\begin{equation}
	\label{eq:Eout_FT_transducer}
	\tilde{\E}_\mathrm{out}(\omega) = \int_{-\infty}^{+\infty}\mathrm{d}\Delta\mathcal{M}(\omega,\Delta)f(\Delta)\,,
\end{equation}
with the integral kernel 
\begin{eqnarray}
	\label{M:omega,Delta}
	\mathcal{M}(\omega,\Delta) &=& \frac{{\rm i}\mu_0L}{\sqrt{2\pi}c}{\rm e}^{{\rm i}\omega L/c'}\int_{0}^{L}{\rm d}z\,{\rm e}^{{\rm i}\delta k z}G(z,\Delta)\\
	& &\times\frac{{\rm e}^{-{\rm i}\omega(L-z)/c_{\rm eff}}{\rm e}^{-d\,h(z,-{\rm i}\omega)}}{{\rm i}(\Delta-\omega)+\gamma/2}\,,\nonumber
\end{eqnarray}
and we have dropped the label $R$, since now all parameters refer to the retrieval dynamics. 

In the following we assume a spatially-independent inhomogeneous broadening, corresponding to $G(z,\Delta)=n(\Delta)/L$. In order to keep the discussion simple, we further neglect $\delta k L$. In this case Eq. \eqref{eq:Eout_FT_transducer} simplifies to
\begin{equation}  
	\label{E_out_FT_simp:omega}
	\tilde{\E}_\mathrm{out}(\omega) = {\rm i}\sqrt{\frac{dn_0L}{c}}\mathcal{C}(\omega){\rm e}^{{\rm i}\omega L/c'}\frac{1-\mathrm{e}^{-{\rm i}2\pi\omega/\omega^{(R)}-d\mathcal{H}(\omega)}}{{\rm i}2\pi\omega/\omega^{(R)}+d\mathcal{H}(\omega)}\,.
\end{equation}
where we have introduced the function
\begin{equation}  
	\label{C:h_omega}
	\mathcal{C}(\omega) = \frac{1}{2\pi n_0}\int_{-\infty}^{+\infty}\mathrm{d}\Delta\frac{n(\Delta)f(\Delta)}{{\rm i}(\Delta-\omega)+\gamma/2}\,.
\end{equation}
When $f$ is independent of the frequency, then $\mathcal C(\omega)\propto\mathcal H(\omega)$, see Eq. \eqref{H:omega}.

For small optical depths, $d \ll 1$, Eq. \eqref{E_out_FT_simp:omega} can be expanded in lowest order in $d$ and reduces to the expression
\begin{equation}  
	\label{E_FT_low_d:w}
	\tilde{\E}_\mathrm{out}(\omega) \approx {\rm i}\sqrt{\frac{dn_0L}{c}}\mathcal{C}(\omega)\mathrm{e}^{{\rm i}\omega(L/c'-\pi/\omega^{(R)})}\mathrm{sinc}\left(\pi\omega/\omega^{(R)}\right)\,,
\end{equation}
where $\mathrm{sinc}(x)=\sin(x)/x$. In this case the frequency $\omega^{(R)}=2\pi c_{\rm eff}/L$ is the upper cutoff of the photon spectral width, and thus gives a lower cutoff to its coherence length.  

\begin{figure}[!ht]
	\centering
	\input{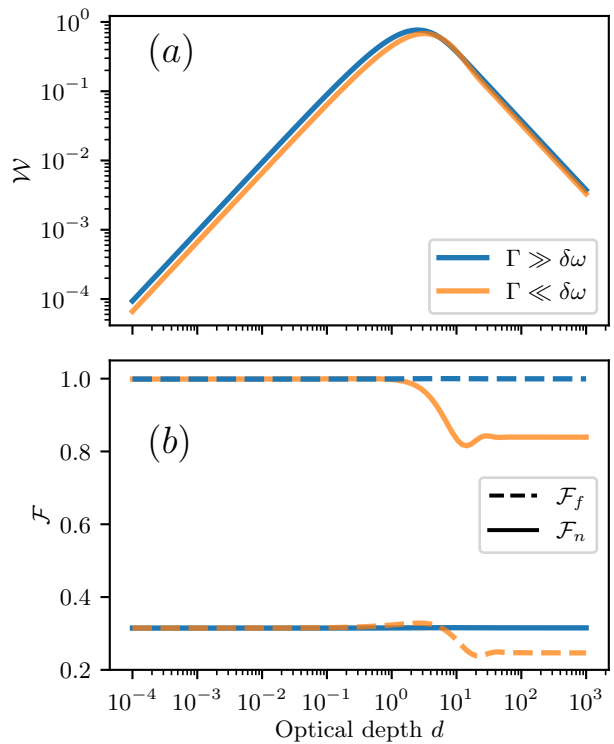}
	\caption{\label{fig:gaussian_photon_generation} (color online) Optical photons from a stored microwave excitation. Subplot (a) displays the retrieval probability $\mathcal W$ as a function of the optical depth $d$. Here, $n(\Delta)=\exp(-\Delta^2/\Gamma^2)/\sqrt{\pi \Gamma^2}$ and $f(\Delta)=\mathcal N {\rm e}^{-\Delta^2/\delta \omega^2}{\rm e}^{{\rm i}\Delta t_\mathrm{c}}$, with $\mathcal N$ being a normalization factor, Eq. \eqref{norm:f}, and $t_\mathrm{c}$ denoting the emission time. Subplot (b) displays the corresponding fidelities $\mathcal F_f$ (dashed lines) and $\mathcal F_n$ (solid lines), Eq. \eqref{eq:fidelity:f}.  The parameters are $\omega^{(R)}=-2\pi\times 30$ GHz, $L=0.01$ m, $\gamma = 2\pi\times 0.01$ GHz and the optical depth $d$ is varied by changing $\mu_0$. The emission time $t_\mathrm{c}$ is $1$ ns and the widths of the Gaussian distributions are $\Gamma/\delta \omega=20$, with $\delta \omega =  2\pi \times 1$ GHz (blue/dark gray line) and $\Gamma/\delta \omega=0.05$, with $\delta \omega = 2\pi\times20$ GHz (orange/light gray line). 
	}
\end{figure}

Let us now assume that the spectral widths of the initial polarization $f$ and of the emitters' distribution $n$ are well below the cutoff $\omega^{(R)}$. We denote the spectral widths of $f$ and $n$ by $\delta \omega$ and $\Gamma$, respectively, with $\delta \omega,\Gamma \ll \omega^{(R)}$. In this limit we can discard the terms proportional to $\omega/\omega^{(R)}$ in Eq. \eqref{E_out_FT_simp:omega} and obtain the expression 
\begin{equation}
	\label{E_FT_large_d:w}
	\tilde{\E}_\mathrm{out}(\omega) \approx {\rm i}\sqrt{\frac{n_0L}{dc}}\frac{\mathcal{C}(\omega)}{\mathcal{H}(\omega)}\left(1-\mathrm{e}^{-d\mathcal{H}(\omega)}\right){\rm e}^{{\rm i}\omega L/c'}\,,
\end{equation}
which is valid for any value of the optical depth. For instance, this expression is consistent with Eq. \eqref{E_FT_low_d:w} for $d\ll 1$ and $\delta\omega,\Gamma\ll\omega^{(R)}$. In this case, it predicts that the emission probability $\mathcal{W}$ scales linearly with $d$. This expresses the collective enhancement of the retrieval process.  For large optical depth, instead, $\mathcal W \sim d^{-1}$ and thus the probability decreases as the optical depth increases, since excitations stored far from the edge cannot propagate through the sample. Figure \ref{fig:gaussian_photon_generation}(a) displays the emission probability, Eq. \eqref{eq:Pemission}, as a function of the optical depth. The emission probability is calculated using Eq. \eqref{E_out_FT_simp:omega} and for Gaussian distributions $f$ and $n$. The two curves correspond to the cases $\delta\omega\ll\Gamma\ll\omega^{(R)}$ and $\Gamma\ll\delta\omega\ll\omega^{(R)}$. In the first case only a small fraction of the emitters are excited, whereas in the latter all emitters are excited with essentially the same amplitude. This, however, has little influence on the retrieval efficiency, which is similar in the two cases.

For the chosen parameters, the emission probability exhibits a peak at optical depths of the order $d\sim 2-3$. The maximal value is $\mathcal W \gtrsim 0.7$. We note that a similar behaviour is also observed in a solid-state quantum memory when the retrieved photon is extracted in the direction of incidence \cite{Afzelius:2009}. 

We now turn to the spectrum of the emitted photon and determine its relation with the initial excitation $f(\omega)$ and with the emitters' spectrum $n(\omega)$ when their spectral widths are much smaller than the cutoff $\omega^{(R)}$. In this case the spectrum can be obtained from Eq. \eqref{E_FT_large_d:w}. We again consider the two limits from above. When the emitters' spectrum is much broader than the one of the stored polarization, $\Gamma \gg \delta \omega$,  we can impose that $n(\Delta)$ is a constant in Eq. \eqref{E_FT_large_d:w}. In this limit $\mathcal H(\omega) \simeq 1/2$, $\mathcal C(\omega) \simeq f(\omega)$ for the functions and parameters considered here and the spectrum of the output photon is the spectrum $f(\omega)$ of the stored polarization to a good approximation. In the opposite case, $\delta \omega \gg \Gamma$, $\mathcal H(\omega)$ and $\mathcal{C}(\omega)$ are functions of $n(\omega)$ and the spectrum thus depends on the emitters' spectrum. For $d \ll 1$, in particular, it takes the simple form $\tilde{\mathcal{E}}_\mathrm{out}(\omega) \propto n(\omega)$. %$\simeq n(\omega)/(2n_0)$

We verify this behaviour by determining the overlap between the field of the emitted photon, Eq. \eqref{E_out_FT_simp:omega}, and the stored polarization, extracted from $\tilde\E_{0,f}(\omega)\propto f(\omega)$ (respectively, of the emitter distribution from $\tilde{\E}_{0,n}\propto n(\omega)$):
\begin{equation}  
	\label{eq:fidelity:f}
	\mathcal{F}_{j=f,n}=\frac{\left|\int_{-\infty}^{+\infty}{\rm d}\omega\tilde{\E}_\mathrm{out}^*(\omega)\tilde{\E}_{0,j}(\omega)\right|}{\sqrt{\int_{-\infty}^{+\infty}{\rm d}\omega''|\tilde{\E}_\mathrm{out}(\omega'')|^2\int_{-\infty}^{+\infty}{\rm d}\omega'|\tilde{\E}_{0,j}(\omega')|^2}}\,.
	%/\mathcal{N}_j\,,
\end{equation}
%with the normalization factor $\mathcal{N}_j=\int_{0}^{+\infty}{\rm d}t'|\E_{0,j}(t')|^2\int_{0}^{+\infty}{\rm d}t''|\Eout(t'')|^2$.
This quantity is maximal and equal to unity when the spectra overlap and when the emission time of the photon corresponds to the time dynamics determined by the phase imprinted on the polarization. 
%
%Moreover, we characterize the spectrum of the emitted photon by calculating its overlap $\mathcal{F}_f$ with a field whose spectral components are $\tilde\E_{0,f}(\omega)\propto f(\omega)$, and thus whose spectrum is the same as the stored polarization. We also calculate the overlap $\mathcal{F}_n$ with a field whose spectral components are $\E_{0,n}(\omega)\propto n(\omega)$, namely, when the spectrum is determined by the emitters' frequency distribution. These are defined as
%\begin{equation}  
%	\label{eq:fidelity:f}
%	\mathcal{F}_{j=f,n}=\left|\int_{0}^{+\infty}{\rm d}t\Eout^*(t)\E_{0,j}(t)\right|^2/\mathcal{N}_j\,,
%\end{equation}
%with the normalization factor $\mathcal{N}_j=\int_{0}^{+\infty}{\rm d}t'|\E_{0,j}(t')|^2\int_{0}^{+\infty}{\rm d}t''|\Eout(t'')|^2$.
%
Figure \ref{fig:gaussian_photon_generation}(b) displays the fidelities $\mathcal F_n$ and $\mathcal F_f$ when the emitters' distribution and the stored excitation have Gaussian spectra and for two opposite limits: $\Gamma \gg \delta \omega$ and $\Gamma \ll \delta \omega$, with the same parameters of the corresponding curves in subplot (a). In the first case, when the spectrum of the stored excitation is narrow, the fidelity $\mathcal F_f$ is close to unity for all values of the optical depth: the spectrum of the emitted photon is given by the spectrum of the stored excitation. When instead the spectral width of the emitters' distribution is narrower than the width of $f(\omega)$, we observe that $\mathcal F_n$ is close to unity only for $d\lesssim 3$ and drops below 0.9 when $d$ exceeds this value. This is a consequence of frequency-dependent absorption significantly changing the shape of retrieved light at large optical depths. The figure thus reflects a change in the shape of the outgoing wavepacket for this situation. For low optical depths the shape directly reflects the initial stored state, but for high optical depth it crosses over to a shape set by a compromise between the initial state and the damping. This analysis suggests that the retrieval probability of the microwave-to-optical transducer depends mainly on the optical depth and seems to be relatively independent on whether the stored excitation has a narrower or broader spectrum than the emitters' distribution. 

\begin{figure}[h]
	\centering
	\input{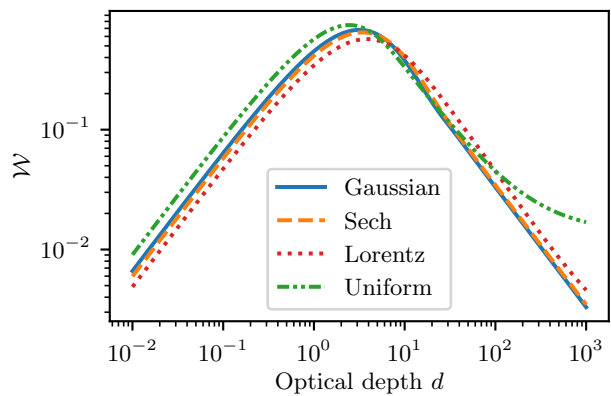}
	\caption{\label{fig:3} (color online) Probability of emission $\mathcal W$ as a function of the optical depth for different frequency distributions $n(\Delta)$. The four lines refer to: (blue) a Gaussian profile $n(\Delta) =\mathcal N\exp(-\Delta^2/\Gamma^2)$; (orange) a sech-profile $n(\Delta)=\mathcal N\mathrm{sech}(\sqrt{\pi}\Delta/\Gamma)$; (red) a Lorentzian profile $n(\Delta) =\mathcal N/(\Delta^2+(\Gamma/\sqrt{\pi})^2)$; (green) a rectangular ("uniform") distribution $n(\Delta)=\mathcal N\theta(\sqrt{\pi}\Gamma/2-\Delta)\theta(\sqrt{\pi}\Gamma/2+\Delta)$, with $\theta(x)$ the Heaviside function and $\mathcal N$ the corresponding normalization, see Eq. \eqref{eq:normalization_cond_G}. The function $f(\Delta)$ and the other parameters are the same as for the case $\Gamma\ll \delta\omega$ of Fig. \ref{fig:gaussian_photon_generation}.}
\end{figure}

Figure \ref{fig:3} displays the emission probability for different forms of $n(\Delta)$ as a function of $d$. Different shapes have maxima for slightly different values of $d\sim 1$, but otherwise show similar behaviour. In general, free-space solid-state quantum memories reach presently optical densities of the order of unity \cite{Thiel:2011,Lauritzen:2012,Afzelius:Eu,Nadia:1,Nadia:2}. We note that the step-like ("uniform") distribution reaches the highest efficiency. This result suggests that the ability to tailor the spectral distribution of the emitters can allow one to optimize the retrieval process and at the same time to tailor the frequency distribution of the emitted photon. This spectral shaping is based on the collective emission properties of the medium and is therefore complementary to protocols based on tailored drive fields \cite{Vasilev:2010,Mueller:2017,Morin:2019,Matthiesen:2013,Keller:2004,Farrera:2016,Eisaman:2004,Farrera2016}.
%Typical values (see for instance https://arxiv.org/pdf/1812.10369.pdf, https://doi.org/10.1103/PhysRevB.85.115111) are between $0.2 - 2$.

  %in ref \cite{Sinclair:2017}:Measurements are carried out
%using a 15.7-mm-long Ti4?Ti3?LiNbO3 waveguide
%
%For the exemplary experimental setup, I would suggest one of Michael Afzelius works with Eu:Y2SiO5.
%https://journals.aps.org/pra/abstract/10.1103/PhysRevA.88.022324.
%https://iopscience.iop.org/article/10.1088/0953-4075/45/12/124001/meta
%
%The crystal in the paper is 1cm long, however, in a single-path configuration, it has optical depth of \apha L = 1.2. If taking a crystal 0.1 mm thick, the optical depth will then be 0.12. If I understand correctly, the dimensionless parameter d is the optical depth \alpha L. This would mean that for a smaller dimension of such a crystal will correspond to the system in your calculations.
%
%Out of our papers, we measured an optical echo in Er:LiYF4 in transmission:
%https://iopscience.iop.org/article/10.1088/1367-2630/aaa7e4
%and the EIT (this would also be an example of optical depth below 1 for a 5mm long crystal in free-space transmission):
%https://arxiv.org/abs/2005.01024

\section{Conclusion} 
\label{Sec:Conclusion and Outlook}

We have analysed the dynamics and efficiency of photon retrieval from solid-state quantum memories acting as quantum transducer. We have focused on the so-called fast protocol, where storage (retrieval) is realised by a $\pi$-pulse, transferring the population to (from) a metastable state and determined the retrieved field as a function of the input photon. We have determined the retrieval dynamics and efficiency taking into account that the emitting and absorbing transitions can be characterized by different internal states, transition frequencies, and emitters' frequency distributions, as is the case for a solid-state quantum transducer. We have discussed the effects which reduce the efficiency of retrieval. 

%The asymmetry between storage and retrieval dynamics limits the retrieval efficiency and gives rise to an optimal optical depth, above which the retrieval probability decreases. One further possible detrimental aspect is when the group velocity of the write pulse and of the read pulse is different from the one of the incident and emitted photon, respectively. This effect corresponds to an effective cutoff frequency which filters out the tails of the spectrum of the retrieved photon. 

Our model can be applied to transducers bridging quite different frequency regimes and to storage and retrieval protocols where the group velocity of the read and write pulses can be different from the one of the photonic excitation. We have exemplary discussed the case of microwave-to-optical frequency conversion and analysed in particular the spectral properties of the emitted photon as a function of the effective cutoff frequency and of the emitters' spectral distribution, which may or may not be correlated between the optical and microwave regime.

One specific limitation to the efficiency of a solid-state quantum transducer in free space is the asymmetry between storage and retrieval dynamics, such that the latter substantially differs from the time-reversal of the storage dynamics. This has generally a detrimental effect, which limits the constructive interference dynamics of photon emission by the individual emitters and can be interpreted as an effective dephasing mechanism. This could potentially be remedied by modulating the read/write pulses using optimal control \cite{Rojan:2014,Koch:2019, Gorshkov:2008:OCT, Giannelli:2018} for the purpose of compensating the accumulated phases. 

\acknowledgements
    
The authors thank Susanne Blum, Pavel Bushev, Stephan Ritter, and especially Nadezhda Kukharchyk for stimulating discussions and helpful comments. GM and TS acknowledge support from the  Deutsche  Forschungsgemeinschaft (DFG, German Research Foundation) Project-ID No.429529648, TRR 306 QuCoLiMa (Quantum Cooperativity of Light and Matter) and SPP 1929 GiRyd (Giant interactions in Rydberg Systems). LG acknowledges funding from the University of Catania, Piano di Incentivi per la Ricerca di Ateneo 2020/2022, proposal Q-ICT. AS gratefully acknowledges financial support from Danmarks Grundforskningsfond (DNRF 139, Hy-Q Center for Hybrid Quantum Networks).
	
\appendix
\section{\label{App:A}Basic equations}

In this appendix we provide some details about the model. We focus on the dynamics of the two-level transition $|g\rangle\to |e\rangle$ and omit the index $(S),(R)$. Starting from the Heisenberg equation of motion, we review the basic steps that lead to the equations for the complex amplitudes reported in Eqs. \eqref{alleqs:eoms_cf}. We describe the $N$ emitters by two-level systems at the positions $z_j$ and transition frequency $\omega_{eg,j}=\weg+\Delta_j$, where $\weg$ is the average frequency, with $j=1,\ldots,N$. We denote the raising operator by $\op{}{\sigma}{_{eg}^{(j)}}=\vk{}{e}{_j}\vb{}{g}{}$. 

The modes of the electromagnetic field are assumed to propagate along the $z$-axis. We distinguish between forward- and backward-propagating modes depending on whether the wave vector points in the positive or negative $z$-direction, respectively. We denote by $\op{}{a}{^\dagger_f}(k)$ and $\op{}{a}{_f}(k)$ ($\op{}{a}{^\dagger_b}(k)$ and $\op{}{a}{_b}(k)$) the operators which create and annihilate, respectively, a forward (backward)-propagating photon at wave number $k$. We assume the linear dispersion relation
\begin{equation}
	\omega(k) = \weg + c(k-\keg)\,,\label{eq:dispersion_relation}
\end{equation}
where $c$ is the group velocity and $\keg$ is the wave number at the transition frequency $\weg$. The operators fulfil the commutation relations $[\op{}{a}{_n}(k),\op{}{a}{^\dagger_m}(k')] = \delta_{nm}\delta(k - k')$, where $\delta(x)$ is the Dirac-delta function, $\delta_{nm}$ is the Kronecker delta, and $n,m \in \{f,b\}$. 

The Hamiltonian describing the quantum emitters, the electromagnetic field modes, and their mutual interaction is given by the operator sum $\op{}{H}{}=\op{}{H}{_0}+\op{}{V}{_f}+\op{}{V}{_b}$, with \cite{Loudon:2000}
\begin{subequations}
	\label{alleqs:Hamiltonian}
	\begin{eqnarray}
	\op{}{H}{_0} &=& \int_{0}^{+\infty}{\rm d}k\hbar \omega(k) \left\{\op{}{a}{^\dagger_f}(k)\op{}{a}{_f}(k)+\op{}{a}{^\dagger_b}(k)\op{}{a}{_b}(k)\right\}\nonumber\\
	&&+ \sum_{j=1}^N\hbar(\weg+\Delta_j)\op{}{\sigma}{_{ee}^{(j)}}\,,\label{eq:H_0}\\
	\op{}{V}{_f} &=& -\hbar g\sqrt{\frac{L}{2\pi}}\sum_{j=1}^N\int_{0}^{+\infty}{\rm d}k\left\{\op{}{\sigma}{_{eg}^{(j)}}\op{}{a}{_f}(k)\mathrm{e}^{{\rm i}kz_j}+\Hc\right\}\,,\nonumber\\
	\op{}{V}{_b} &=& -\hbar g\sqrt{\frac{L}{2\pi}}\sum_{j=1}^N\int_{0}^{+\infty}{\rm d}k\left\{\op{}{\sigma}{_{eg}^{(j)}}\op{}{a}{_b}(k)\mathrm{e}^{-{\rm i}kz_j}+\Hc\right\}\,,\nonumber\\
	\label{eq:V}
	\end{eqnarray}
\end{subequations}
where $g$ is the coupling constant and $L$ is the medium's length in $z$-direction. The interactions $\op{}{V}{_f}$ and $\op{}{V}{_b}$ are here assumed to be in the electric-dipole and rotating-wave approximation. 

The photon field in position space is given in the Heisenberg picture by the slowly-varying annihilation operators \cite{Gorshkov:2007,Blum:2013}
\begin{eqnarray}
\op{}{\E}{_f}(z,t) &=& \sqrt{\frac{L}{2\pi}}\mathrm{e}^{i\weg t}\int_{0}^{+\infty}{\rm d}k\op{}{a}{_f}(k,t)\mathrm{e}^{i(k-\keg)z}\,,\nonumber\\\
\op{}{\E}{_b}(z,t) &=& \sqrt{\frac{L}{2\pi}}\mathrm{e}^{i\weg t}\int_{0}^{+\infty}{\rm d}k\op{}{a}{_b}(k,t)\mathrm{e}^{-i(k-\keg)z}\,.\nonumber\\\label{eq:def_photon-field_annihilation_operator}
\end{eqnarray}
To describe the dynamics of the emitters, we resort to a coarse graining by dividing the $z$-axis into a grid of finite steps $\delta z$ and the frequency range into a grid with finite steps $\delta\Delta$.  We define the normalized distribution of emitters
\begin{equation*}
G(z,\Delta)=\frac{1}{N}\sum_{j=1}^N\delta(\Delta-\Delta_j)\delta(z-z_j)\,.
\end{equation*}
The number of emitters at the position $z_j\in[z-\delta z/2,z+\delta z/2]$ and with detuning $\Delta_j\in[\Delta-\delta\Delta/2,\Delta+\delta\Delta/2]$ is given by:
$$
	N_{z,\Delta}=G(z,\Delta)\delta z\delta\Delta\,.
$$ %denotes the number of emitters whose positions are in the interval $z_j\in[z-\rm{d}z/2,z+\rm{d}z/2]$ and with detuning $\Delta_j\in[\Delta-\rm{d}\Delta/2,\Delta+\rm{d}\Delta/2]$. 
The polarization operator can be written as \cite{Kroell:2001,Iakoupov:2013}
\begin{equation}
\label{eq:def_polarization_operator}
\op{}{P}{}(z,t;\Delta) = \frac{\sqrt{N}}{N_{z,\Delta}}\sum_{j\in S_{z,\Delta}}\op{}{\sigma}{_{ge}^{(j)}}(t){\rm e}^{{\rm i}\weg t}\,,
\end{equation}
where $S_{z,\Delta}$ includes all atoms with $z_j\in[z-\delta z/2,z+\delta z/2]$ and $\Delta_j\in[\Delta-\delta\Delta/2,\Delta+\delta\Delta/2]$.

For single photons almost all of the atoms remain in the ground state $\ket{g}$. The commutation relations of the coarse-grained operators then read
\begin{subequations}
	\label{eqs:commutation_relations}
	\begin{eqnarray}
	\left[\op{}{\mathcal{E}}{_n}(z,t),\op{}{\mathcal{E}}{^\dagger_m}(z',t)\right] &=& L\delta_{nm}\delta(z-z')\,,\label{eq:commutation_relation_E}\\
	\left[\op{}{P}{}(z,t;\Delta),\op{}{P}{^\dagger}(z',t;\Delta')\right] &=& \frac{1}{G(z,\Delta)}\delta(z - z')\delta(\Delta - \Delta')\,,\nonumber\\
	\label{eq:commutation_relation_P}
	\end{eqnarray}
\end{subequations}
where the Dirac-delta function shall be understood in terms of this coarse graining, see also Ref. \cite{Blum:2013} for details. We remark that the commutation relation \eqref{eq:commutation_relation_P} is defined for $G(z,\Delta)\neq 0$.%, otherwise the polarization operator vanishes. In fact, when $G(z,\Delta)=0$, then at the corresponding tuple $(z,\Delta)$ there are no emitters and the polarization operator is identically zero. 

The Heisenberg equations of motion for the fields and polarization are \cite{Gorshkov:2007}
\begin{subequations}
	\label{alleqs:eoms}
	\begin{eqnarray}
	\left(\frac{\partial}{\partial t} + c \frac{\partial}{\partial z}\right)\op{}{\E}{_f} &=& {\rm i}\gN L\int_{-\infty}^{+\infty}{\rm d}\Delta G(z,\Delta)\op{}{P}{}(z,t;\Delta)\mathrm{e}^{-{\rm i}\keg z}\,,\nonumber\\
	\label{eq:eom_E}\\
	\left(\frac{\partial}{\partial t} - c \frac{\partial}{\partial z}\right)\op{}{\E}{_b} &=& {\rm i}\gN L\int_{-\infty}^{+\infty}{\rm d}\Delta G(z,\Delta)\op{}{P}{}(z,t;\Delta){\rm e}^{{\rm i}\keg z}\,,\nonumber\\
	\label{eq:eom_E:b}\\
	\frac{\partial}{\partial t}\op{}{P}{} &=&-(\gamma/2  + {\rm i} \Delta) \op{}{P}{}+\hat\eta(z,t;\Delta)\nonumber\\
	&&+i\gN\left(\op{}{\E}{_f}\mathrm{e}^{{\rm i} \keg z}+\op{}{\E}{_b}\mathrm{e}^{-{\rm i} \keg z}\right)\,.
	\label{eq:eom_P}
	\end{eqnarray}
\end{subequations}
Here $\gamma$ is the polarization decay rate and $\op{}{\eta}{}(z,t;\Delta)$ is the corresponding Langevin force with a vanishing mean value $\langle \hat\eta(z,t;\Delta)\rangle=0$ and the only non-vanishing second-order moment being $\langle \hat\eta(z,t;\Delta)\hat\eta^\dagger(z',t';\Delta') \rangle \propto \delta(z-z')\delta(\Delta-\Delta')\delta(t-t')$ \cite{Hald:2001}.

By means of the operators $\op{}{\E}{_f}$, $\op{}{\E}{_b}$, and $\op{}{P}{}$ it is possible to write a generic state of the system with a single excitation:
\begin{widetext}
	\begin{equation}
	\vk{}{\Psi(t)}{} = \frac{1}{\sqrt{cL}}\left(\sum_{n\in\{f,b\}}\int_{-\infty}^{+\infty}{\rm d}z\E_n(z,t)\op{}{\E}{^\dagger_n}(z,0)+L\int_{0}^{L}{\rm d}z\int_{-\infty}^{+\infty}{\rm d}\Delta G(z,\Delta)P(z,t;\Delta)\op{}{P}{^\dagger}(z,0;\Delta)\right)\vk{}{g,\dots,g;\mathrm{vac}}{}\label{eq:generic_state_of_system}\,.
	\end{equation}
\end{widetext}
Here, $\vk{}{g,\dots,g;\mathrm{vac}}{}$ denotes the state with all emitters in $\ket{g}$ and the electromagnetic field modes in the vacuum state $\ket{\mathrm{vac}}$ and $\E_f$, $\E_b$, and $P$ are the amplitudes for the excitation (photon) to be respectively forward and backward propagating or in an atomic state at position $z$ and time $t$. We note that in principle the full state of the system should contain terms describing photons emitted into other modes through spontaneous emission, corresponding to the noise operator $\hat{\eta}$. Since such photons are lost from the system they are of no interest to us here and we omit these terms for simplicity. This means that our state is not fully normalized, but the amplitude in the forward or backward direction still gives the correct retrieval efficiency, which is our main interest. The equations of motion for $\E_f$, $\E_b$, and $P$ are obtained from the Heisenberg equations of motion \eqref{alleqs:eoms} by taking the matrix element, e.g., $\E_f(z,t) = \sqrt{\frac{c}{L}}\vb{}{g,\dots,g;\mathrm{vac}}{}\op{}{\E}{_f}(z,0)\vk{}{\Psi(t)}{}$, and are given in Eqs. \eqref{alleqs:eoms_cf}. 

The temporal shape of the photon leaving the medium at $z=L$ is given by the expectation value \cite{Gorshkov:2007}
\begin{equation}
\label{eq:def_temporal_shape_output_field}
\mathcal I(t) = \frac{c}{L}\vb{}{\Psi(t)}{}\op{}{\mathcal{E}}{^\dagger_f}(L,0)\op{}{\mathcal{E}}{_f}(L,0)\vk{}{\Psi(t)}{}
\end{equation}
and can be now expressed in terms of the complex amplitude $\mathcal{E}_f$ as $\mathcal{I}(t) = |\mathcal{E}_f(L,t)|^2$ (as long as one keeps in mind that this replacement is valid solely for nomally ordered expressions \cite{Gorshkov:2007PRL}). 	

\section{\label{App:SpinWave}Fast storage and retrieval in a third metastable level}

In this appendix we discuss the details of the classical pulses used for the fast storage and retrieval. We denote by $S(z,t;\Delta)$ the spin wave, which is coupled by a laser pulse to the polarization $P^{(j)}(z,t;\Delta)$, with $j=S,R$. Assuming that during the pulse the coupling with the photon field can be neglected, the equations determining the dynamics are
\begin{subequations}
	\label{Eqs:pulse}
	\begin{eqnarray}
	\frac{\partial}{\partial t} P^{(j)}(z,t;\Delta)&=& -i\Delta P^{(j)}(z,t;\Delta)\ -{\rm i}\frac{\Omega^{(j)}(z,t)}{2} S(z,t;\Delta)\,,\nonumber\\
	~\\
	\frac{\partial}{\partial t} S(z,t;\Delta)&=& -{\rm i}\frac{\left(\Omega^{(j)}(z,t)\right)^*}{2} P^{(j)}(z,t;\Delta)\,.
	\end{eqnarray}
\end{subequations}
%where we discard the free evolution during the dynamics, taking the pulse duration $\bar t$ such that $\delta\omega \bar t\ll 1$, with $\delta\omega$ the spectral width of the incident photon.
The storage pulse propagates in the backward direction with the incident photon and has an area of $\pi$, ideally implementing perfect population transfer from state $|e\rangle$ to state $|r\rangle$: 
\begin{eqnarray}
\label{Eq:OmegaS}
\Omega^{(S)}(z,t)&=&\Omega_0{\rm e}^{-{\rm i}k'^{(S)}z}\theta(t-t_0(z))\theta(\bar t+t_0(z)-t)\,,\nonumber\\
\end{eqnarray} 
where $\Omega_0=\pi/{\bar t}$ and
$$t_0(z)=t_0+\delta t+(L-z)/c'^{(S)}\,.$$
In the above expression we have introduced the time delay $\delta t$ between the photon and the laser pulse at the position $z=L$. Under the assumption that in the relevant spectral range $|\Delta|\bar{t}\ll 1$ holds, the transfer can be considered to be instantaneous and the spin wave at position $z$ after the pulse reads
\begin{equation}
	S(z,t_0(z)^+;\Delta) = -{\rm i} {\rm e}^{{\rm i}k'^{(S)}z}P^{(S)}(z,t_0(z);\Delta)\label{Eq:Storage:P}
\end{equation}		
and has been imprinted a phase grating. %We note that the time $T_1$ in the exponential function is a constant, $T_1=t_0(z)+z/c'=t_0+\delta t + L/c'$.
At time $T_1 = t_0+\delta t + L/c'^{(S)}$ the control pulse leaves the medium and the transfer to the spin wave is completed everywhere in the medium. %Moreover, in case $c'>c$
We remark that in the case $c'^{(S)}>c^{(S)}$, the delay time shall be chosen in order to preserve the temporal sequence between photon excitation and storage pulse. A sufficient condition is $\delta t\ge L(1/c^{(S)}-1/c'^{(S)})$. Since the photon is typically stored in the region of size $L/d^{(S)}$, a more modest bound is $\delta t\gtrsim L/d^{(S)} (1/c^{(S)}-1/c'^{(S)})$. 

The retrieval dynamics is described by Eqs. \eqref{Eqs:pulse}, now with a pulse propagating in the forward direction and after implementing the map $\Delta\to p[\Delta]$: 
\begin{eqnarray}
\label{Eq:OmegaR}
\Omega^{(R)}(z,t)&=&\Omega_0{\rm e}^{{\rm i}k'^{(R)}z}\theta(t-t_1(z))\theta(\bar t+t_1(z)-t)\,,\nonumber\\
\end{eqnarray} 
with
$$
	t_1(z)=T_1+T_S+z/c'^{(R)}
$$
and $T_S$ is the storage time. We assume that there is no dynamics during the storage time in the spin wave $S$ (which requires that the state $|s\rangle$ is perfectly degenerate along the crystal). In reality there will always be some dynamics in the spin wave leading to decay of the stored excitation, but this is a separate issue and we shall not go into it here. The retrieved polarization at position $z$ takes the form 
\begin{eqnarray}
\label{Eq:P:Retrieval}
P^{(R)}(z,t_1(z)^+;{p[\Delta]})&=&- {\rm i}{\rm e}^{{\rm i}k'^{(R)}z}S(z,t_1(z);\Delta)\,,
\end{eqnarray}		
such that $P^{(R)}(z,t;{p[\Delta]})$ vanishes for $t<t_1(z)^+$.
Inserting Eq. \eqref{Eq:Storage:P} in Eq. \eqref{Eq:P:Retrieval} after using $S(z,t_1(z);\Delta) = S(z,t_0(z)^+;\Delta)$ we obtain the expression of Eq. \eqref{P:R}.

\section{\label{App:different_broadenings} Correlated and uncorrelated inhomogeneous broadenings}
In section \ref{sec:retieval} we have introduced the map $p[\Delta]$ that connects the inhomogeneous broadenings of the storage and retrieval transitions. This allows us to model both correlated and uncorrelated broadenings. In this appendix we give the integral kernel $\mathcal{S}$, Eq. \eqref{eq:F:v,u}, connecting the input and output fields in a more general form and provide examples for both types of broadenings and the related maps $p$.

Let $\mathcal{G}(\Delta,\Delta')$ be the distribution for the detunings $\Delta$ and $\Delta'$ of the storage and retrieval transitions, respectively. The kernel $\mathcal{S}$ in terms of this distribution reads 
\begin{widetext} 
	\begin{eqnarray}
		\mathcal{S}(\omega,\omega') &\equiv& \frac{\sqrt{d^{(S)}d^{(R)}}}{2\pi n_0}\sqrt{\frac{c^{(S)}}{c^{(R)}}} {\rm e}^{{\rm i}\omega(T_1+T_S+L/c'^{(R)})}{\rm e}^{-{\rm i}\omega'(T_1-L/c'^{(S)})}\int_{0}^{L}{\rm d}z\,{\rm e}^{{\rm i}(\delta k^{(S)}+\delta k^{(R)})z}\\
		& &\times \int_{-\infty}^{+\infty}{\rm d}\Delta\int_{-\infty}^{+\infty}{\rm d}\Delta' \,\,\frac{\mathcal{G}(\Delta,\Delta')G^{(R)}(z,\Delta')\,{\rm e}^{-{\rm i}\omega (L-z)/c_{\rm eff}^{(R)}}{\rm e}^{-d^{(R)} h^{(R)}(z,-{\rm i}\omega)}{\rm e}^{-{\rm i}\omega'(L-z)/c_{\rm eff}^{(S)}}{\rm e}^{-d^{(S)} h^{(S)}(z,-{\rm i}\omega')}}{({\rm i}(\Delta'-\omega)+\gamma'/2)({\rm i}(\Delta-\omega')+\gamma/2)}\,.\nonumber
	\end{eqnarray}
\end{widetext}
In the CRIB protocol, for instance, one realizes the distribution $\mathcal{G}(\Delta,\Delta')=\delta(\Delta + \Delta')$. This is an example for correlated broadenings and the resulting kernel $\mathcal{S}$ can be recovered from Eq. \eqref{eq:F:v,u} when choosing $p[\Delta]=-\Delta$. The inhomogeneous broadenings are instead uncorrelated if the distribution is, for instance, of the form $\mathcal{G}(\Delta,\Delta')=\mathcal{G}_0(\Delta')$. In this case, $p$ samples  detunings $\Delta'$ from the distribution $\mathcal{G}_0(\Delta')$.

\end{document}